\documentclass[useAMS,usenatbib]{mn2e}
\usepackage{graphicx}
\usepackage{subfigure}
\usepackage{booktabs}
\usepackage{verbatim}
\usepackage{color}

\def   \aj {{\rm {ApJ}}}
\def   \araa {{\rm {ARA\&A}}}
\def   \apj {{\rm {ApJ}}}

\def   \apjs {{\rm{ApJS}}}

\def   \aap {{\rm{A\&A}}}

\def   \aaps {{\rm{A\&AS}}}

\def   \mnras {{\rm{MNRAS}}}

\title[Spectroscopy and linear spectropolarimetry of the early Herbig Be stars PDS 27 and PDS 37]{Spectroscopy
  and Linear Spectropolarimetry  of the early
  Herbig Be Stars PDS 27 and PDS
  37\thanks{Based on observations collected at the European Southern
    Observatory (ESO), Paranal, Chile under programme ID 084.C-0952(A), 088.C-0858(A), 088.C-0858(B) and 088.C-0858(C).}}

\author[K. M. Ababakr, J. R. Fairlamb, R. D. Oudmaijer and M.E. van den Ancker]{K. M. Ababakr$^{1}$\thanks{E-mail:
pykma@leeds.ac.uk}, J. R. Fairlamb$^{1}$, R. D. Oudmaijer$^{1}$ and M.E. van den Ancker$^{2}$ \\
$^{1}$School of Physics and Astronomy, University of Leeds, EC Stoner Building, Leeds LS2 9JT, UK\\
$^{2}$European Southern Observatory, Karl-Schwarzschild-Str.2, D-85748 Garching bei M\"{u}nchen, Germany}

\begin{document}

\date{Accepted 2015 July 2.  Received 2015 July 1; in original form 2014 September 29}

\pagerange{\pageref{firstpage}--\pageref{lastpage}} \pubyear{2002}

\maketitle

\label{firstpage}

\begin{abstract}

The number of well-studied early-type pre-main-sequence objects is
very limited, hampering the study of massive star formation from an
observational point of view. Here, we present the results of VLT/FORS2
spectropolarimetric and VLT/X-shooter spectroscopic observations of
two recently recognised candidate Herbig Be stars, PDS 27 and PDS 37. Through
analysis of spectral lines and photometry, we find that these two objects are hot,
17500 $\pm$ 3500~K, have large radii, 17.0 $\pm$ 4.0 and 25.8 $\pm$ 5.0~${\rm R}_\odot$, and are very
massive, 15.3 (+5.4, -4.4) and 21.1 (+11.0, -5.3)~M$_{\odot}$ for PDS 27 and PDS 37, respectively. This
suggests that these two objects are very young in their evolution and
may become O-type stars. Their youth is supported by their high
accretion rates of the order of $10^{-3}$--$10^{-4.5}~{\rm M}_\odot$/yr. A
change in linear polarisation across the absorption component of
H$\alpha$ is detected in both objects. This change indicates that the
circumstellar environment close to the star, at scales of several stellar radii, has a flattened structure, which we identify
as an inner accretion disc. Strong variability is seen in both
objects in many lines as further indication of an active circumstellar
environment.
\end{abstract}


\section{Introduction}

Herbig Ae/Be (HAeBe) stars are optically visible pre-main-sequence
(PMS) stars with masses roughly between 2 and
10~M$_{\odot}$. With their intermediate masses, they bridge the gap
between low-mass stars whose formation is well understood and high-mass stars whose formation is poorly understood
\citep{2007ARA&A..45..481Z}. HAeBe objects were first
identified by \citet{1960ApJS....4..337H} as stars of spectral type A
or B with emission lines and which illuminate a bright nebula in their
surroundings. Their spectral energy distributions (SEDs) are characterised by
an infrared excess due to the dust in the circumstellar environment
\citep{1998ARA&A..36..233W}.

HAeBe objects not only have some common characteristics with high-mass
objects such as clustering \citep{1999A&A...342..515T} but also have
spectropolarimetric characteristics in common with T Tauri stars
\citep{2002MNRAS.337..356V}. Therefore, the study of the formation of
HAeBe stars can bring out the differences in the mechanisms of the
formation of low- and high-mass stars. In addition, because they are
closer and less embedded than massive young stars, HAeBe stars can act as a
powerful test bed for the formation of massive stars.

To study the star formation mechanism as a function of mass, it is
necessary to sample a large number of HAeBe stars, with many
representatives in all mass bins. The known sample of HAeBe stars [see
e.g. the catalogue by \citet{1994A&AS..104..315T}] contains many
A-type and late B-type objects, but there is a dearth of the earliest
type, most massive Herbig Be (HBe) stars (also see
e.g. \citet{2011A&A...536L...1O}). In this paper we present an
in-depth investigation of two early HBe star candidates identified by
the Pico dos Dias Survey (PDS). Based on their observational characteristics, PDS 27 (also known as DW CMa, RA 07:19:36,
Dec. $-$17:39:18) and PDS 37 (also known as Hen 3-373, RA 10:10:00,
Dec. $-$57:02:07.3) were proposed to be early type Herbig stars by \citet{2003AJ....126.2971V}. The young nature of PDS 27 was also suggested by \citet{2006A&A...458..173S}. Both stars were also classified as young in the Red MSX Source (RMS) survey for Massive Young Stellar Objects (MYSOs) \citep{2013ApJS..208...11L}. In 2011, \citet{2011A&A...526A..24V} considered PDS 27 to be an evolved star instead, but ruled out an evolved nature for PDS 37. However, as we will demonstrate in this paper, both objects are very similar, and we proceed under the assumption that they are young stars.  

Using optical photometry, the temperature has been estimated to be around 22000 K for both objects \citep{2003AJ....126.2971V}. \citet{2003AJ....126.2971V} also estimated their distance and it was found to be 1100 pc for PDS 27 and 720 pc for PDS 37; PDS 27 is associated with CMa star-forming region while PDS 37 is associated with C 282.4+0. \citet{2013MNRAS.429.2960I} studied the near-infrared (NIR) first
overtone CO emission of PDS 37, and assuming that it originates from
the circumstellar disc, they found the disc to be highly inclined to
the line of sight ($80^{\circ}$). \citet{2009ApJ...698.2031R}
determined an intrinsic polarisation of 2.77\% for PDS 37, suggesting
that the circumstellar environment is not spherical. 

Here we present a spectroscopic and spectropolarimetric study of these
two objects. We will use the large wavelength coverage provided by
X-shooter to determine the stellar parameters, evolutionary status and
accretion rates. Spectropolarimetry is a very powerful technique that
can probe very small unresolved spatial scales of the order of stellar radii
and allows one to extract information on the geometry of this
material (see e.g. \citet{1999MNRAS.305..166O,2002MNRAS.337..356V,2005MNRAS.359.1049V}).
A new element in the present study is that the medium-resolution spectropolarimetry has a wavelength coverage from 4560 to 9480 $\rm \AA$, which is much larger than presented in any previous
study at similar spectral resolution.

This paper is organised as follows: in Section 2 the observations and
data reduction are discussed. The results are presented in Section
3. Section 4 presents an analysis of the data followed by a discussion
in Section 5. Finally, conclusions are drawn in Section 6.

\section{Observations}

\subsection{Spectropolarimetry}

The data of PDS 27 and PDS 37 were taken as part of a wider
spectropolarimetric investigation whose results will be presented in a
future paper (Abakakr et al., in preparation). The targets were
selected based on their strong emission lines, allowing an
investigation into their spectropolarimetric behaviour. 

The spectropolarimetric data for our targets were obtained during the
nights of 2012 February 4 and 5 with the FORS2 spectrograph mounted on
ESO's VLT in Chile. Three gratings were used with a $4096\times2048$
pixel CCD to sample the entire optical range. The 1400V grating with
central wavelength 5200~$\rm\AA$ covered the spectral range from
4560 to 5860~$\rm\AA$ and includes H$\beta$.  The 1200R grating was used for the spectral range from
5750 to 7310~$\rm\AA$, with central wavelength 6500~$\rm\AA$,
and includes notable lines: H$\alpha$, [O I] 6300~$\rm\AA$, and He~I
lines at 5875, 6678 and 7065~${\rm \AA}$. Finally, the 1027Z grating
covers $7730$ -- $9480~{\rm \AA}$, at central
wavelength 8600~$\rm\AA$, and this includes the calcium triplet lines
and the hydrogen Paschen series.  A 0.5 arcsec slit was used in all
cases, which resulted in a spectral resolution of R = 3800, 4800, 4800
in the {\it V}, {\it R} and {\it Z} bands respectively.

FORS2 is equipped with an optical polarimeter in order to analyse the
linearly polarised components. The polarisation optics consist of a
rotating half-wave plate and a calcite block. The calcite block
separates the incoming light into two perpendicularly polarised light
beams, the ordinary (O) and extraordinary (E) beam. The half-wave
plate is rotated to measure polarisation at four different angles:
$0^{\circ}$, $45^{\circ}$, $22.5^{\circ}$, and $67.5^{\circ}$. One
complete set of observations consists of four exposures, one at each
of the four angles. To avoid saturation in the bright emission lines,
several sets of observations were taken with short exposure times.

The polarisation accuracy is determined only by photon statistics; in
order to reach a relative accuracy of 0.1\%, one needs to detect
roughly one million photons per resolution element. The targets are
relatively faint, with {\it V}-band magnitudes of 13.0 and 13.5 for
PDS 27 and PDS 37 respectively \citep{2003AJ....126.2971V}, so the
total integration times were comparatively long with 1.6 and 2.4h
total. For PDS 27 the exposure times are broken down into 8$\times$340s for
the V-grating, 24$\times$50s for the R-grating and 16$\times$120s for the
Z-grating. Similarly for PDS 37 the times are 16$\times$340s for the
V-grating, 24$\times$70s for the R-grating and 16$\times$100s for the
Z-grating. Polarised and unpolarised standard stars were observed
throughout the night. The exposure times for these were much shorter
as it was only the continuum values that were of interest.

The spectropolarimetric data reduction was carried out using {\sc gasgano}
and FORS pipeline v4.9.18. We have used two recipes: one dedicated to the calibration and the other to
extract the science spectrum. The reduction procedure consists of bias
subtraction, flat fielding, extraction and wavelength calibration of
the O and E spectra. The Stokes parameters, obtained from the O and E
beams, lead to the percentage of linear polarisation, {\it P}, and
polarisation angle, $\theta$, according to the following equations:


\begin{equation}
 P = \sqrt[]{Q^{2}+U^{2}}
\end{equation}	

\begin{equation}
\theta = \frac{1}{2} \arctan \bigg(\frac{U}{Q}\bigg)	
\end{equation}


Finally, the spectropolarimetric data were imported into the {\sc iraf} and
{\sc polmap} package for measurement and analysis.

FORS2 is mounted on the UT1 Cassegrain focus, resulting in a
relatively stable instrumental polarisation. To improve signal-to-noise and the polarisation accuracy, the individual data sets were
combined. The instrumental polarisation is determined from the
observation of unpolarised standard stars. This was found to be
$\sim$ 0.16\%. The polarisation angle has an absolute error of $\sim$
$0.5^{\circ}$ as determined from observing polarised standard stars. We did not
perform a correction for instrumental polarisation as it is small
compared to the continuum values of the targets, while our main
emphasis lies  on the differential polarisation between lines and
continuum.

\begin{figure*}
        \centering
        \includegraphics[width=16cm]{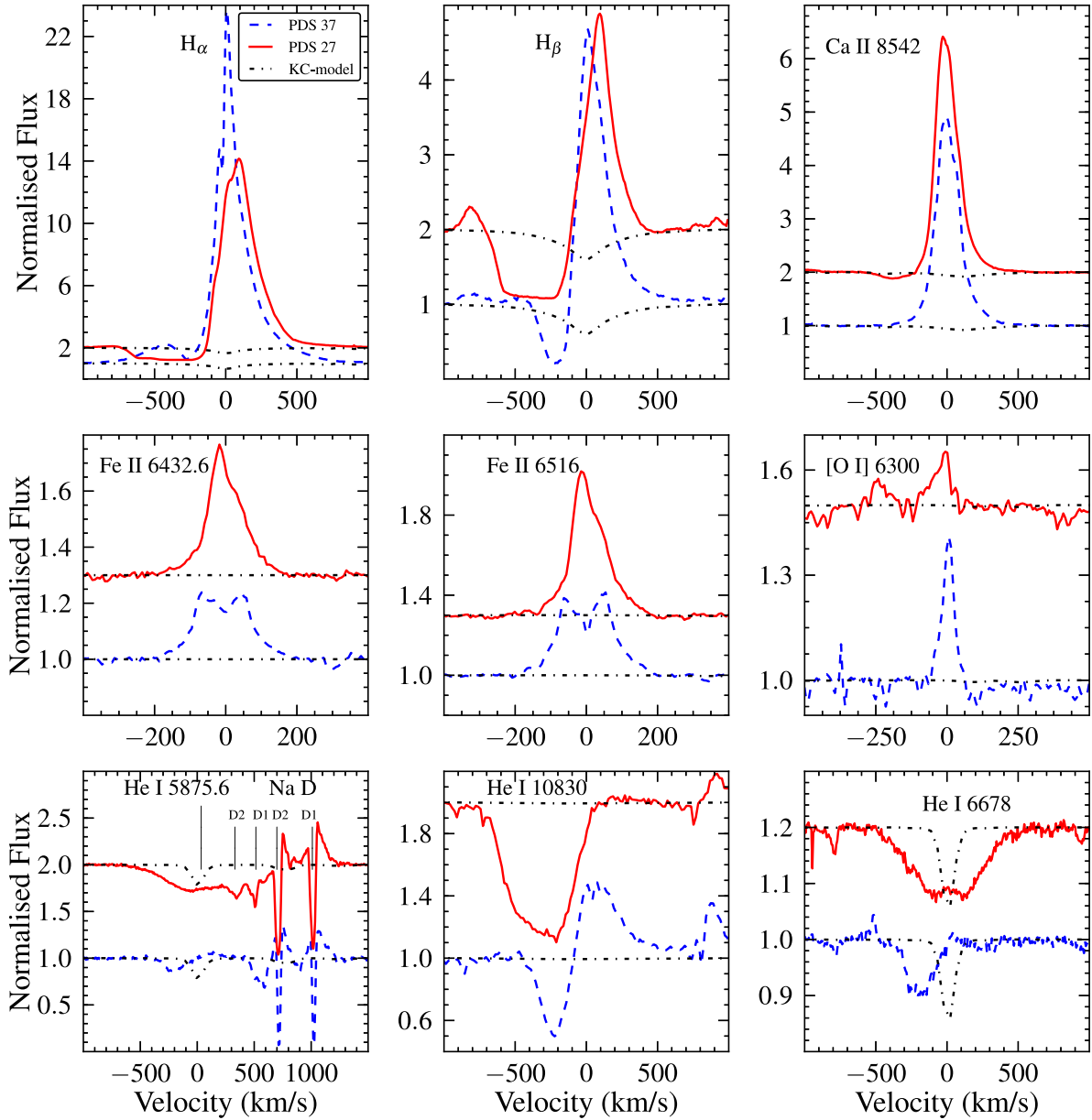}
   \caption{A selection of observed spectral lines from the X-shooter
     data is shown for both targets (for PDS 27 only the X-shooter(b) data set is shown
     for clarity). The spectra are shifted such that their systemic
     velocity (see the text) is 0 km$\rm s^{-1}$. The red solid line denotes
     PDS 27, while the blue dashed line is PDS 37. The black dash-dotted line is a
     Kurucz-Castelli (KC) model atmosphere for a 21000~K type star
     \citep{Kurucz93,2004astro.ph..5087C}; this is an upper limit to the temperature (this limit will be discussed in Section 4.1.3). The spectra of both objects are normalised to one but PDS 27 is shifted by an arbitrary value for clarity. Note the similarity of both objects'
     spectra, with a difference being the number of double-peaked
     lines of Fe II in PDS 37. Blueshifted absorption reaching -700 and -400 km$\rm s^{-1}$ is observed in H$\alpha$
     and helium for PDS 27 and PDS 37, respectively.}
\label{allline}
\end{figure*}

\subsection{Spectroscopy}

Spectra of PDS 27 and PDS 37 were obtained using the medium-resolution
spectrograph X-shooter, mounted on the VLT, Chile
\citep{2011A&A...536A.105V} as part of a larger programme
(\citealt{2011AN....332..238O, Fairlamb15}). One of
the main strengths of the instrument is its huge wavelength coverage, taken simultaneously, over three arms: 3000 - 5900$\rm
\AA$, UVB; 5300$\rm \AA$ - 1.0${\rm \mu}$m, VIS; and 1.0 - 2.4${\rm
  \mu}$m, NIR.  The smallest slit widths available (0.5 arcsec - UVB, 0.4 arcsec
- VIS and 0.4 arcsec - NIR) were used and provided the highest spectral
resolution available for X-shooter: {\it R}=9100, 17400 and 11300,
respectively. Both objects were observed in nodding mode in an ABBA
sequence. Two observations were made of PDS 27, on the evenings of 2009
December 18 and 2010 February 24 (these data sets will be referred to
as X-shooter(a) and X-shooter(b), respectively, for the remainder of
this paper). The exposures were the same for each observation with a
breakdown across the arms of UVB - 300s$\times$4, VIS -
300s$\times$4 and NIR - (2s$\times$6)$\times$4. PDS 37 was observed
on the evening of 2010 March 31 with exposures of UVB - 300s$\times$4,
VIS 300s$\times$4, and NIR (3s$\times$6)$\times$4. The spectra of both
objects were reduced using the X-shooter pipeline v0.9.7
\citep{Modigliani2010} following standard reduction procedures
(flat-fields, wavelength calibration, etc.). The final wavelength calibration was carried out manually by identifying several telluric absorption lines using a solar spectrum catalogue \citep{Delbouille73}.

The UVB is corrected for instrumental response across
the Balmer jump region (in order to measure the Balmer excess, which is
a diagnostic for the strength of the accretion). To do this the observed spectra
of the flux standards were divided through by the known flux from
CALSPEC\footnote{http://www.stsci.edu/hst/observatory/crds/calspec.html}. The
resulting response curve was then applied to the target of that night.
To measure the Balmer excess, we do not need an absolute flux
calibration, only the correct shape of the SED is needed. This method of calibration is therefore
sufficient. We require only knowledge of the line strengths, profile
shapes and calibrated wavelengths in the other arms. Therefore, this calibration is not needed in the other arms.

\section{Results}

\subsection{Spectral lines}

The spectra of both stars are remarkably similar showing numerous
emission lines, while there is a paucity of
photospheric absorption lines. A representative set of lines is shown
in Fig. \ref{allline}.

\begin{figure}
	\centering
	\includegraphics[width=1\columnwidth]{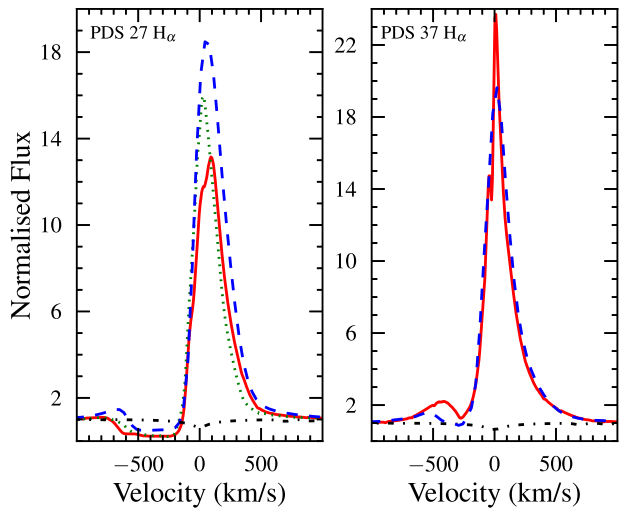}

        \caption{H$\alpha$ line variability from three observation
          epochs for PDS 27: FORS2 (blue dashed line) and two X-Shooter
          epochs, X-shooter(a) (green dotted line) and X-shooter(b) (red solid
          line).  For PDS 37, there is only one X-shooter data set (red solid
          line). A KC model (black dash-dotted line) of temperature 21000~K
          is also overplotted. The lines are shifted to the rest
          wavelength of the stars as in Fig.~1.}
        \label{hvari}	

\end{figure}

The strongest emission is found among the hydrogen recombination
lines, Fe II, Ca II and [O I]. The H$\alpha$ and H$\beta$ lines have
P Cygni profiles, with absorption observed at velocities in excess of
500 kms$^{-1}$ in the case of PDS 27, with some emission observed bluewards
of the P Cygni absorption. The H$\alpha$ lines are very strong with
line-to-continuum ratios of the order of 20 and an equivalent width (EW) of
-70.4 and -111.4 $\rm\AA$ for PDS 27 and PDS 37, respectively. Several
strong Fe II emission lines can be seen in the visible portion of the
spectra; the majority of these are double peaked in PDS 37, while they
are all single peaked in PDS 27. Strong emission is also present in the Ca II triplet (8498, 8542,
8662~$\rm \AA$) and Ca II doublet (8912, 8927~$\rm \AA$) in both stars. The Ca II triplet lines are blended with the
Paschen lines at 8502, 8545 and 8665~$\rm\AA$. The
Ca II line at 8542 and 8662~$\rm\AA$ in PDS 27 has a very weak
blueshifted absorption. The other nearby Paschen lines at
8598, 8750 and 8863~$\rm\AA$ have similar absorption
components. Therefore, the absorption seen in
the Ca II lines is more likely to belong to the Paschen series with which they are
blended. He I lines at 5875, 6678, 7065 and
10830~$\rm\AA$ are in absorption and blueshifted to the same
velocities as observed in the hydrogen P Cygni profiles. He I at
10830~$\rm\AA$ has a P~Cygni profile in PDS 37 with absorption
observed up to -450 kms$^{-1}$, and the EW of the emission is slightly
larger than that of the absorption. For PDS 27, He I at 10830~$\rm\AA$
has a very strong absorption up to -700 kms$^{-1}$ and is accompanied
by a very weak redshifted emission component. As we go from low-order lines in the hydrogen Balmer series
(H$\alpha$, H$\beta$) towards the higher order hydrogen lines, we see
a progressively weaker absorption component in each line. The only
exception is that the absorption in H$\alpha$ is less deep than in
H$\beta$, which we shall discuss later. The varying hydrogen line
profiles prevent a reliable measurement of the stars' radial velocity
({\it V$_{\rm r}$}). Therefore, to obtain the systemic velocity, we measure the
{\it V$_{\rm r}$} from the clean and symmetric Fe II and Ca II triplet lines in the X-shooter and FORS2 spectra. The measured {\it V$_{\rm r}$} are corrected for
heliocentric motion and local standard of rest motion for the dates of
observation. These give local standard of rest velocities ({\it V$_{\rm LSR}$}) of $47.2\pm3.4$, $45.3\pm4.3$, and $63\pm7$~kms$^{-1}$ for PDS 27 X-shooter(a), X-shooter(b) and FORS2, respectively. Both X-shooter(a) and X-shooter(b) values are in agreement with each other, while there is a difference of $\sim18\pm8$~kms$^{-1}$ between X-shooter and FORS2. For PDS 37, the {\it V$_{\rm LSR}$} is found to be $11.7\pm5.5$~kms$^{-1}$ for X-shooter and $13.5\pm8.0$~kms$^{-1}$ for FORS2. For the final {\it V$_{\rm LSR}$} to PDS 37 an average of the X-shooter and FORS2 values is taken and this results in a {\it V$_{\rm LSR}$} of $12.6\pm4.8$~kms$^{-1}$.

Both stars' spectra are variable. An illustration of this is provided
in Fig. \ref{hvari} in which we show H$\alpha$ at three observation
epochs for PDS 27 and two for PDS 37. The line profiles are variable
with a deeper absorption component when the emission is weaker.  For
PDS 27 the H$\alpha$ line-to-continuum ratio ranges from 13 to 18, and the
EW increases from $-70.4$ to $-120.8~\rm\AA$ from the
X-shooter(b) to the FORS2 data sets (with the X-shooter(a) observations lying in between). For PDS 37 the line-to-continuum
increases from 19.5 to 24, and the EW from -111.4 to
$-122.6~\rm\AA$. \citet{2011A&A...526A..24V} observed, in 1990, an EW
of -88 and $-105~\rm\AA$ for PDS 27 and PDS 37 respectively.

\begin{figure}
        \centering
        \includegraphics[width=1\columnwidth]{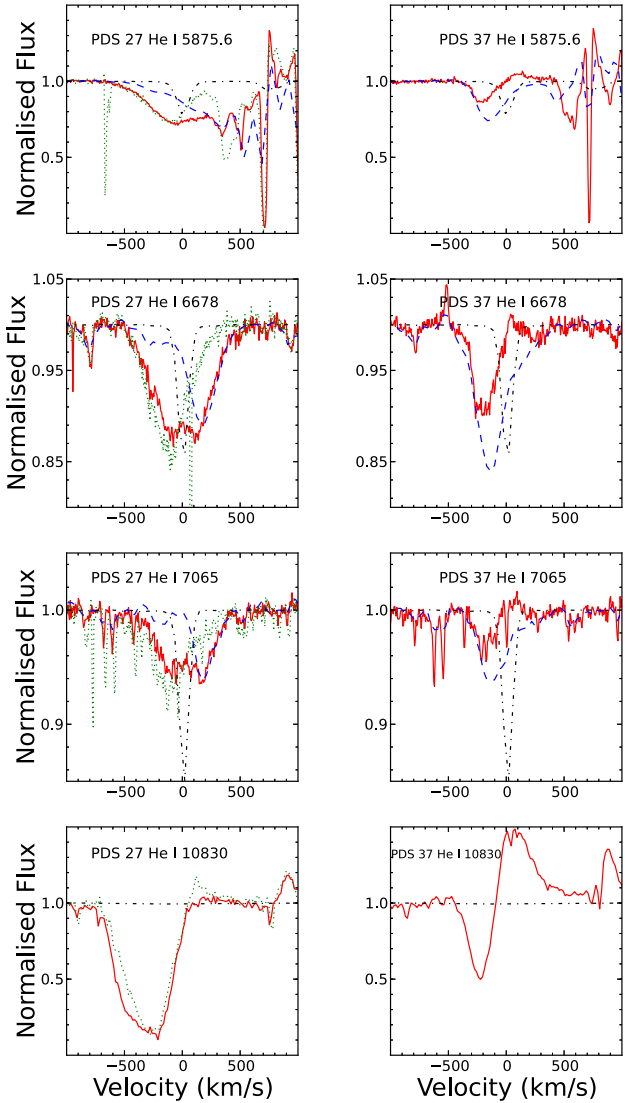}

        \caption{He I line variability is shown for PDS 27 of three observation
          epochs: FORS2 (blue dashed line) and two X-shooter
          epochs, X-shooter(a) (green dotted line) and X-shooter(b) (red solid
          line).  For PDS 37, there is only one X-shooter data set (red solid
          line) and one FORS2 observation (blue dashed line). A KC model (black dash-dotted line) of temperature 21000~K
          is overplotted for the 5875, 6678 and 7065~$\rm \AA$ lines,
          while for the 10830~$\rm \AA$ line a black dash-dotted line shows
          just the continuum level. The lines are shifted to the rest
          wavelength of the stars as in Fig.~1.}

        \label{hevari}    
\end{figure}

Variability of the He I line in PDS 27 also occurs between all
observations. Fig. \ref{hevari} shows four of the He I lines, along with
overplots from each observation epoch in order to highlight the
variability. The He I lines 5875, 6678 and 7065~$\rm\AA$ are present
in both X-shooter and FORS2 spectra, while 10830~$\rm\AA$ is covered only by the
X-shooter observations. The 5875~$\rm\AA$ line in the FORS2 data shows
a broad blue absorption wing which is seen extending from -500
kms$^{-1}$ towards the red, but we cannot quantify how far it extends into the red
due to the Na I lines blending with it. In the X-shooter(a) data set a
clear absorption line is seen centred at $\sim$-150 kms$^{-1}$ from the rest
wavelength. The line in the X-shooter(b) is difficult to measure, due
to the blending mentioned, but it appears to be centred at $\sim$-100
kms$^{-1}$. The 6678 and 7065~$\rm\AA$ lines show a similar
behaviour in their velocities between the data sets, as demonstrated in
Fig. \ref{hevari}. The 10830~$\rm\AA$ line only shows a small
variability in the strength of the line.

In the case of PDS 37, all the He I lines, present in both X-shooter
and FORS2, show variability in the line centre of $\sim$50~kms$^{-1}$
between them, with respective line centres being blueshifted from the
central wavelength. Some of these show a very weak redshifted emission,
which is seen to be much stronger in the 10830~$\rm \AA$ line. There is also strong
variability in the line strengths, measured solely across the
absorption component, where the strength is $\sim$2.5 greater in FORS2
than in X-shooter.

\begin{table}
\centering
\caption{The observed EW and the {\it E(B-V)} values obtained from the DIBs. Errors in the EW and {\it E(B-V)} are typically 5\%. }
\label{dib}
\begin{tabular}{c c c   c c c}
\hline
DIB&EW/{\it E(B-V)}&\multicolumn{2}{c}{PDS 27}&\multicolumn{2}{c}{PDS 37}\\
    &  & EW             &  {\it E(B-V)}   &     EW      &   {\it E(B-V)}\\
($\rm\AA$)&(m$\rm\AA$/mag.)&(m$\rm\AA$)&   (mag)    &(m$\rm\AA$) &   (mag) \\
\hline
5780.6   &579&     571.5             &     0.99    &     278.9               & 0.48   \\
5797.1   &132&   190.7             &    1.44   &    221.6              & 1.68   \\
5849.8   &48&    116.4            &    2.42   &     139.4              & 2.90   \\
6089.8   &17&    24.1           &    1.42   &     23.4              & 1.38   \\
6204.3   &189&   230.8           &   1.22  &     129.0           & 0.68   \\
6013.7   & 23&  242.0         &    1.05  &     206.5          & 0.89   \\
6660.6   &51&   61.8        &     1.21 &      46.9           & 0.90   \\
6993.2   & 116&   131.7          &   1.06  &    100.0           & 0.86  \\
\hline
\end{tabular}
\end{table}

\subsection{Interstellar lines}

The spectra of both targets contain interstellar absorption lines,
most notably the K I doublet (7664.9, 7698.9~$\rm \AA$), Na I doublet
(5889.9, 5895.9~$\rm \AA$) and the diffuse interstellar bands (DIBs);
their origin comes from the material present in the interstellar
medium (ISM). The strength of the DIBs increases with the amount of
material, and therefore reddening, along the line of sight
\citep{1994A&AS..106...39J}. However, the DIBs can only provide a
lower limit to the total reddening. \citet{1997MNRAS.291..797O} found for a sample of young
objects that the majority of the DIBs correlate with the reddening due to interstellar
dust, but they remain insensitive to the dust in circumstellar
envelopes and their parental molecular clouds. Eight well-known DIBs are clearly identified in both objects;
the measured EW for each line, along with the line strength per unit
{\it E(B-V)} determined from \citet{1994A&AS..106...39J}, and the
resulting {\it E(B-V)} are given in Table \ref{dib}. The overall {\it
  E(B-V)} is found to be $1.2 \pm 0.16$ for PDS 27 and $0.98 \pm 0.37$
for PDS 37. The spread of the values is fairly large with {\it E(B-V)}
ranging, for PDS 27 and 37 respectively, from a low of 0.99 and 0.48, for the
5780~$\rm\AA$ line, to a high of 2.42 and 2.9, for the 5849~$\rm\AA$ line. The very
large colour excess derived from the 5849~$\rm\AA$ DIB is not entirely
unexpected as it was suggested by \citet{1986ApJ...305..455C} that the
strength of this DIB could provide a good indication of the total
reddening towards an object including the circumstellar component. This
argument is strengthened by \citet{1997MNRAS.291..797O} who found in
their study that this DIB was an exceptionally case which more accurately traced the total reddening
for each star than any of the other DIBs. Therefore, in the overall
{\it E(B-V)} reported above, the 5849~$\rm\AA$ DIB is not included. The {\it V$_{\rm LSR}$} of the DIBs is measured to be $26.6\pm5.3$, $27.3\pm5.6$ and $27.5\pm2$~kms$^{-1}$ for PDS 27 X-shooter(a), X-shooter(b) and FORS2 spectra, respectively; all the values are in very good agreement with each other. For PDS 37, the {\it V$_{\rm LSR}$} is found to be $0.5\pm2.8$ and $6\pm4$~kms$^{-1}$ for the X-shooter and FORS2 spectra, respectively. This is consistent with
the {\it V$_{\rm LSR}$} observed in the sharp absorption lines from Na I
5889.9 and 5895.9~$\rm\AA$, which is found to be $26.5\pm1$, $26.8\pm0.3$ and $26\pm2$~kms$^{-1}$ for PDS 27 X-shooter(a), X-shooter(b) and FORS2 spectra, respectively, and $-4.5\pm0.5$ and $-5\pm1$ for PDS 37 X-shooter and FORS2 spectra, respectively.

\begin{table*}
 \centering

\caption{The continuum polarisation of both objects measured in the
  following wavelength regions: {\it B} band centred at 4700~$\rm
  \AA$, {\it V} band centred at 5500~$\rm \AA$, {\it R} band centred
  at 7000~$\rm \AA$, {\it I} band centred at 9000~$\rm \AA$. The
  polarisation was measured over a wavelength range of 250~$\rm \AA$
  either side of the central wavelength, except for the {\it B} band
  where it was 125~$\rm \AA$ either side. All the errors are differential, where the systematic error is $\sim$0.16\% in polarisation and $\sim$
$0.5^{\circ}$ in angle (see the text in Section 2.1 for details).}

\label{co}
\begin{tabular}{c c c c c c c c c}
\hline
Object&\multicolumn{2}{c}{{\it B}}&\multicolumn{2}{c}{{\it V}}&\multicolumn{2}{c}{{\it R}}&\multicolumn{2}{c}{{\it I}}\\
& $P_{cont}(\%)$&$\theta_{cont}^{\circ}$&$P_{cont}(\%)$&$\theta_{cont}^{\circ}$&$P_{cont}(\%)$&$\theta_{cont}^{\circ}$&$P_{cont}(\%)$&$\theta_{cont}^{\circ}$\\
\hline 
PDS 27   &  8.60~$\pm$~0.03  &  18.64~$\pm$~0.11   &   9.00~$\pm$~0.01  &   18.87~$\pm$~0.01   &   8.83~$\pm$~0.01  &   18.36~$\pm$~0.04   &  7.49~$\pm$~0.01  &  18.12~$\pm$~0.03 \\
PDS 37   &  4.68~$\pm$~0.04  &  129.97~$\pm$~0.27  &   5.12~$\pm$~0.01  &   130.80~$\pm$~0.01  &   5.18~$\pm$~0.01  &   129.98~$\pm$~0.06  &  4.38~$\pm$~0.01  &  133.38~$\pm$~0.05 \\
\hline
\end{tabular}
\end{table*}

\subsection{Continuum polarisation}

\begin{figure}
	\centering
	\includegraphics[width=1\columnwidth]{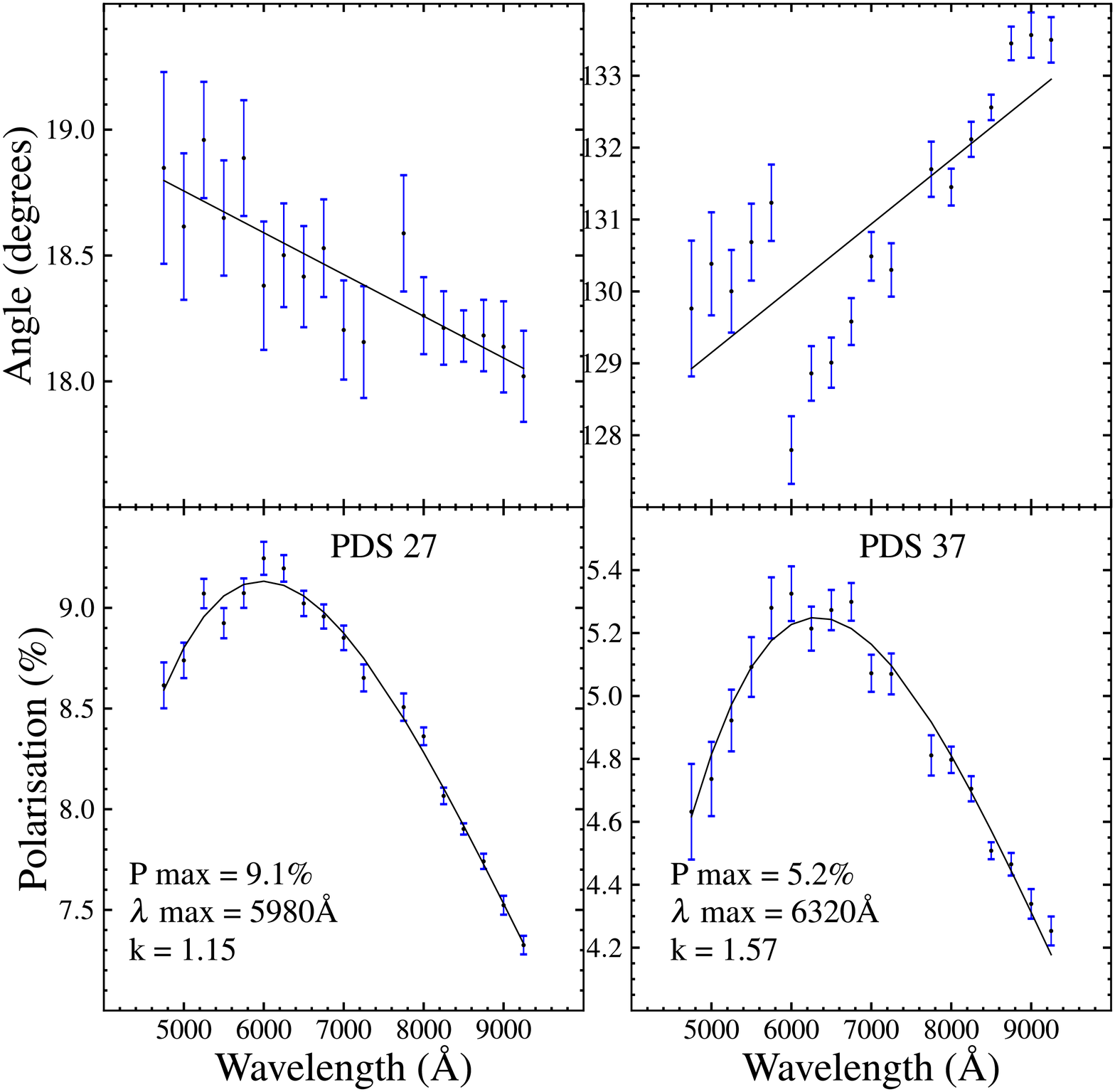}

        \caption{The polarisation and polarisation angles as a
          function of wavelength for PDS 27 and PDS 37. The
          spectropolarimetric data were coarsely sampled to improve
          the error bars. The black lines denote the best-fitting
          Serkowski law to the data.}

	\label{ser}
\end{figure}

The continuum polarisation of the targets is shown in Fig.
\ref{ser}. In the figure, the spectropolarimetric data were rebinned
using a coarse sampling, of $\sim 250~{\rm \AA}$ per bin, to minimize errors.  It can be seen that the
percentage of polarisation increases from short wavelengths, peaks in
the {\it V} band and then decreases towards longer wavelengths.

The observed continuum polarisation is a combination of interstellar
polarisation and polarisation due to circumstellar dust and electron
scattering, if the dust and electrons deviate from circular symmetry. The nature of interstellar polarisation is well described and
explained by \citet{1975ApJ...196..261S}. According to Serkowski's
law, the linear polarisation follows an empirical curve according to
the following equation:

\begin{equation}
 \frac {P(\lambda)} {P_{max}} = {\rm exp}\bigg[-k~{\rm ln}^{2} \bigg(\frac{\lambda_{max}}{\lambda}\bigg)\bigg]
\end{equation}
where $\lambda_{max}$ is the wavelength where the polarisation is at its
maximum value, $P_{max}$, and {\it k} is the width of the empirical curve. Typically, $\lambda_{max}$ is $\sim$ 5500~$\rm\AA$,
but can be in a range of 4500-8000~$\rm\AA$.

The Serkowski law provides an excellent fit to the
data and is shown in Fig. \ref{ser}. This combined with the large {\it E(B-V)} found towards these objects in Section 3.2 suggests that the bulk of the observed polarisation is indeed  
interstellar. The angle of polarisation changes slightly over the
different bands; a change of 1 deg is observed for PDS 27 while a 5
deg change is observed for PDS 37. The polarisation angle
normally does not change as a function of wavelength for the ISM. The
slight change in position angle with wavelength indicates that there
is a contribution from another polarising agent to the total
polarisation. The results of continuum polarisation are summarised in
Table \ref{co}.

\subsection{Line spectropolarimetry}

\begin{figure*}
        \centering
        \includegraphics[width=16cm]{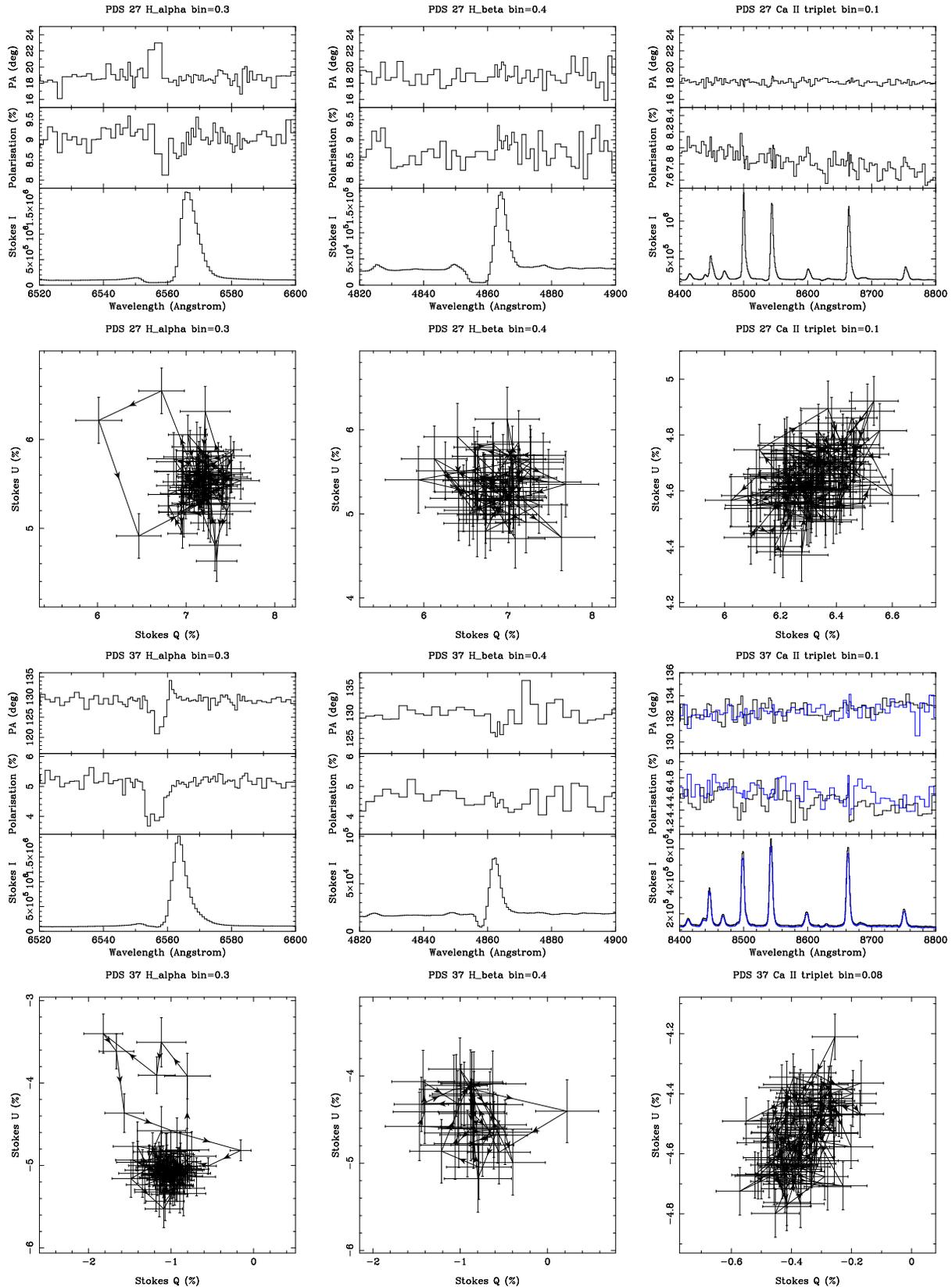}
        
   \caption{The polarisation data of PDS 27 (upper half) and PDS 37
     (lower half) are shown here. The data are presented as a
     combination of triplots (top) and ({\it Q,U}) diagrams (bottom). In the
     triplot polarisation spectra, the Stokes intensity ({\it I}) is shown in
     the lowest panel, polarisation (\%) in the centre, while the
     position angle (PA) is shown in the upper panel. The {\it Q} and {\it U}
     Stokes parameters are plotted against each other below each
     triplot. The data are rebinned to a constant value, as indicated
     at the top of each plot. The Ca II triplet in PDS 37 is presented in two colours: the first two data sets with possible line effect are shown in black solid line; the second two data sets with no line effect are shown in blue solid line.}
     \label{pds}
\end{figure*}

The polarisation spectra around the strongest lines, H$\alpha$,
H$\beta$ and the calcium triplet are displayed in Fig. \ref{pds}. There is a clear change in polarisation across the H$\alpha$ line in
both objects. The red peak and, where visible, the blue peak have
almost the same polarisation value as the continuum, while there is a
strong change across the absorptive component in both polarisation
spectra and the ({\it Q,U}) plots. The shift in the polarisation from
continuum to the line is of the order of 1\% for PDS 27 and 1.6\% for PDS 37.

Fig. \ref{pds} also shows that there is no clear change in the
polarisation spectra across H$\beta$ for both targets. Several
different ways of rebinning the data were tried, but no compelling
evidence for line effects across H$\beta$ could be found. Our
explanation for this will be discussed in Section 5.1.

The Ca II triplet has a lower ionisation potential than the H$\alpha$
line, and as such it originates in a different region at a larger
distance from the central stars. We did not detect any line effect across the Ca II triplet in
PDS 27, whereas in PDS 37, the first two data sets show a possible line effect while the second two data sets do not show such a line effect (see Fig. \ref{pds}). Therefore, the current
data are not sufficient to confirm whether there is a line effect or
not across the Ca II triplet in PDS 37. We do not detect changes across any other lines in this polarimetric
spectrum, which includes the Ca II doublet, He~I, [O~I] 6300~$\rm\AA$
and Fe II lines.

The observed line characteristics of the targets in the FORS2 data are
presented in Table \ref{li}, including an indication whether a line
effect is present.

\begin{table*}
 \centering
 \begin{minipage}{140mm}
\caption{ The line spectropolarimetry results: columns (2), (3), (5) and (6) list the Stokes ({\it I}) characteristics; columns (4) and (7) list line spectropolarimetry characteristics of each target. Errors in the EW measurements are typically 5\%.}
\label{li}
\begin{tabular}{l r r c r r c }
\hline
Lines      &   \multicolumn{3}{c}{PDS 27}  &   \multicolumn{3}{c}{PDS 37} \\
           &  EW~($\rm\AA$)  &  Line/cont.  &  Line effect  &  EW~($\rm\AA$)  &  Line/cont.  &  Line effect \\
\hline 
H$\alpha$  &     -120.80   &     18.0   &     Yes       & -122.60     &    19.4     &    Yes        \\
H$\beta$   &     -14.56   &       5.8  &      No        &  -12.27     &     4.5     &    No         \\
Ca II(8498)&     -24.05   &       6.0  &     No       & -22.66     &   4.4      &    --          \\
Ca II(8542)&     -23.97   &      5.3   &      No       & -25.43     &   5.1       &     --         \\
Ca II(8662)&    -22.78   &      5.2   &      No      & -23.12     &     4.8     &      --        \\
Ca II(8912)&    -1.11   &       1.2  &     No        & -0.81    &     1.1     &     No         \\
Ca II(8927)&    -1.79   &      1.3  &     No        & -1.69       &   1.3       &     No         \\
He I(6678) &     0.90    &   0.9   &     No        & 1.40      &    0.8      &     No         \\
He I(7065) &     0.27   &       0.9  &     No        & 0.52       &   0.9       &     No         \\
O I(6300)  &    -0.36    &      1.1   &     No        & -0.46     &   1.2       &     No         \\
Fe II(6456)&    -3.14   &      1.8  &     No        & -2.07      &   1.5       &     No         \\
\hline
\end{tabular}
\end{minipage}
\end{table*}

\section{Analysis}

\subsection{Stellar parameters}

\subsubsection{Stellar velocity}

Due to the lack of photospheric absorption lines in the spectra of both stars, we used the clean and symmetric Fe II and Ca II triplet emission lines to estimate the systemic velocity.

For PDS 27 we noted that in Section 3.1 both X-shooter observations provide a similar {\it V$_{\rm LSR}$}, while there is a difference of $\sim18\pm8$~kms$^{-1}$ in {\it V$_{\rm LSR}$} between X-shooter and FORS2. To examine the reliability of this variation, it is important to check the {\it V$_{\rm LSR}$} of the interstellar lines as this should not vary over different epochs. In Section 3.2, we found a similar {\it V$_{\rm LSR}$} of the DIBs in all epochs. These velocities are consistent with the {\it V$_{\rm LSR}$} observed in the sharp absorption lines from Na I 5889.9 and 5895.9~$\rm\AA$. The observed variation in {\it V$_{\rm LSR}$} of PDS 27 could be due to the star itself through changes in the circumstellar environment. On the other hand, it could be due to the presence of a second star in the system. Distinguishing conclusively between the two scenarios is challenging with only a few epochs of data, and requires additional observations. However, we found that the line profiles are almost symmetric in all epochs; hence, it is more likely that the changes seen are due to binary motions and are not due to the movements of circumstellar material around the star. Based on the above argument, we decide not to use our stellar velocities for PDS 27 for kinematic distance determinations.

For PDS 37, the {\it V$_{\rm LSR}$} is consistent within the errorbars for the X-shooter and FORS2 data.

\subsubsection{Distance}

Accurate stellar parameters are required for any quantitative analysis
to be made about either age, accretion or evolution for these two
enigmatic objects. First we will discuss the distance, which will later help
constrain the total luminosity.

Using a Galactic rotation curve an estimate of the distance to the
stars can be made using the observed velocities. In this work, we use
the rotation curve from \citet{Reid09}. The observed {\it V$_{\rm LSR}$} in PDS 27 is unlikely to trace the actual stellar velocity as argued in the previous section. Fortunately, molecular observations of
$^{13}$CO(1-0) and NH3(1,1) towards these objects are available
\citep{Urquhart07,Urquhart11}. The molecular lines in this case trace the parental clouds in a larger volume around
the star compared to the optical lines. These authors observe velocities
of 43.5 and 5.4 kms$^{-1}$ for PDS 27 and PDS 37, respectively. 
Using the molecular velocities the resulting kinematic distance for PDS 27 is 3.17 (+0.66, -0.62) kpc, while for PDS 37 the distance is found to be 3.67 $\pm$ 0.95~kpc. For PDS 37 the stellar lines can also be used; these provide distances of 4.53 $\pm$ 0.77 and 4.74 (+0.75, -0.90) kpc for the lines measured from X-shooter and FORS2, respectively. The molecular and spectral line measurements are in agreement with each other in PDS 37. For the final distance to PDS 37, an average of the three values is taken and this results in a distance of 4.31 $\pm$ 0.67 kpc. 

An interesting consistency check is to see what the kinematic
distances to the DIBs along the line of sight towards the objects are. This is because the DIBs trace the material between us and the star, and should therefore be closer than the stellar and molecular distances.
The observed {\it V$_{\rm LSR}$} velocities of the DIBs in Section
3.2 provide kinematic distances for PDS 27 of 1.76 (+0.58, -0.55) and 1.80 (+0.58, -0.55) kpc for the X-shooter and FORS2 data, respectively. For PDS 37 the kinematic distances of DIBs are found to be 2.59 $\pm$ 1.4 and 3.76 $\pm$ 0.92 kpc for the X-shooter and FORS2 data, respectively. These distances
are indeed lower than the ones traced by the stellar and molecular lines, which is as expected.

\subsubsection{Temperature}

Spectral types of both stars are difficult to determine due to
a general lack of photospheric absorption lines in the spectra. So far, the only spectral type given to these stars is by \citet{2003AJ....126.2971V}, who assign them both a tentative B2 classification based on their photometry.

A rough lower limit on the temperature can be found using the observed colour indices. \citet{2003AJ....126.2971V} measure a $(B-V)_{obs}$ of 1.32 for PDS 27 and 1.52 for PDS 37. The DIBs provide colour excesses, $E(B-V)$, of $1.20 \pm 0.16 $ and $ 0.98 \pm 0.37 $ for PDS 27 and PDS 37, respectively. Combining the observed and excess colours allows an intrinsic $(B-V)$ to be derived. This is only a lower limit as the DIBs trace the ISM, but not necessarily the total extinction. The highest $(B-V)_{int}$ from the DIBs is found to be -0.04 in PDS 27 and 0.15 in PDS 37. This gives a lower limit to the temperatures which corresponds to B9-type star with a temperature of 10000~K and A5-type star with a temperature of 8000~K for PDS 27 and PDS 37, respectively. The stars are likely to be even hotter than this, as this is a lower limit.

An upper limit to the temperature can be placed using the Balmer jump in the spectra.
This is because the Balmer jump decreases in size with increasing temperature (for temperatures greater than $\sim$~9000~K, where the Balmer jump is at its maximum). The observed size of the jump therefore places an upper limit on the photospheric temperature. Moreover, a small jump can also be achieved by cooler stars combined with a flux excess due to accretion. The Balmer jump constraint provides an upper limit of 21000~K. A discussion of the jump and possibility of an excess flux will be given in Section 4.2.1.

The DIBs and Balmer jump constraints give an effective temperature range of 8000~K/10000 - 21000~K. In addition, the observed lines from the two objects suggest that they lie more towards the upper temperature limit.

 We detect N~I emission lines at 8629 and 8683~$\rm\AA$ in both objects. These lines have a high excitation energy ($\sim$10~eV), and have only been observed in B-type HAeBe stars, predominately early types \citep{hamann92}.  Given that the latest spectral type in which these authors detected both NI emission lines was B6 (corresponding to 14000 K), we infer a revised lower limit to the temperatures of PDS 27 and PDS 37 of around 14000 K.

Based on the above three arguments, which combine knowledge of photometry and spectra, we find the most plausible range to the temperatures to be 14000 -- 21000 K.  We choose to adopt a temperature of 17500~$\pm$~3500~K for both objects.

In addition, the only absorption lines that are present in both spectra are from He I. If their origin were assumed to be from the photosphere, then their strength would peak at around 21000~K. However, the observed strengths are greater than predicted from stellar model atmospheres, and they are asymmetric. The blueshifted absorption extends all the way to -500 kms$^{-1}$  and covers the same velocity range as the hydrogen lines. Therefore, an additional contribution to the absorption could be due to a strong wind. Although strong absorption has been regularly observed for  T Tauri stars in the He I 10830~$\rm \AA$ line, such a strong and broad absorption in a wind is not expected in the higher energy transitions of He I in classical T Tauri stars (CTTs) by either theory \citep{2011MNRAS.416.2623K} or observations \citep{Beristain2001}. This suggests that the stars are much hotter than CTTs, which is in agreement with the other temperature constraints adopted.


\begin{figure*}
        \centering
        \subfigure{
                \includegraphics[width=8cm]{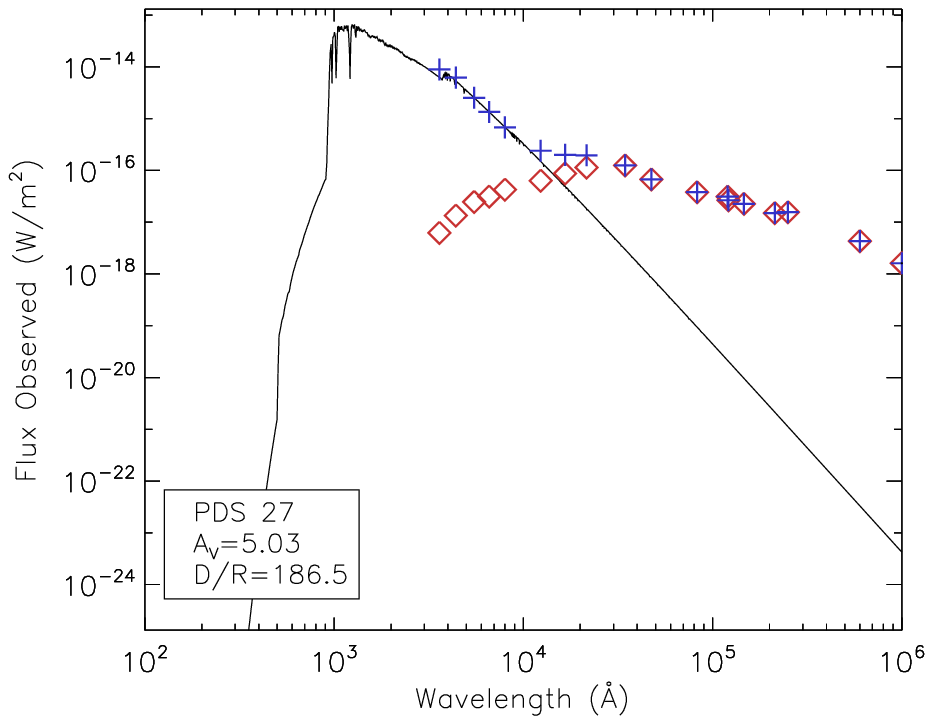}}      
         \subfigure{
                \includegraphics[width=8cm]{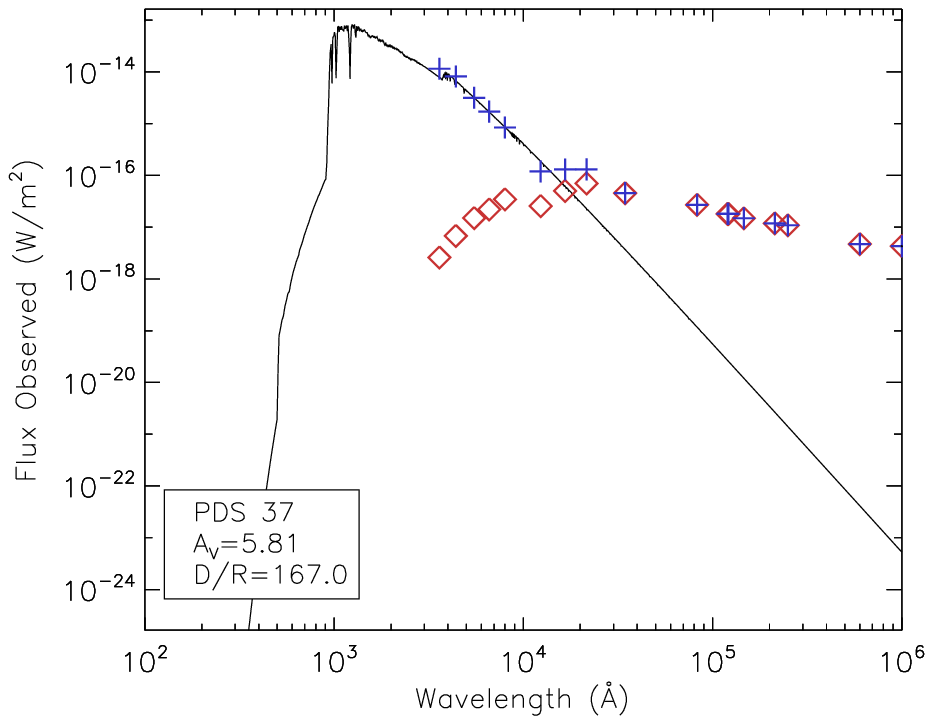}}
          \caption{An example of the SED integration for extracting a
            dereddened total flux for the case of PDS 27 (left) and
            PDS 37 (right). The red diamonds are the observed
            photometry, the blue crosses are the dereddened
            photometry and the solid black line is a KC model of
            17500~K.}
   \label{fig:l_extrapolate}
       \end{figure*}

\subsubsection{Luminosity}

With a limit on both distance and temperature established for the stars,
other stellar parameters can now be found. A start will be made with
determining the luminosity using the previously observed optical photometry. A total
flux needs to be obtained, and due to the limited amount of photometric
observations an extrapolation into the UV must be performed
to do this. To assess the UV contribution, Kurucz-Castelli (KC) models are used which
provide a surface flux density. The {\it BVRI} photometry is then dereddened with this law until their shape matches the slope of the KC model for the adopted temperature.
The {\it U} band is excluded due to possible effects of an excess due to accretion, while the longer wavelength photometry is also excluded due
to being affected by an IR excess from the dust. To fit the models to the
photometry, a scaling factor must be applied, as the models used are in surface flux while the photometry is observed flux. This scaling factor is just a ratio between
the distance and radius squared, ({\it D/R})$^2$. For a temperature of 17500  $\pm$ 3500~K,
a respective $A_V$ and {\it D/R} are found to be 5.03 $\pm$ 0.13 and 186.5 $\pm$ 21.0 for PDS 27, and 5.81 $\pm$ 0.13 and 167.0 $\pm$ 18.9 for PDS 37. These
reddening values are consistent with the DIB derived values, which
were lower limits to the total reddening.
Fig. \ref{fig:l_extrapolate} shows the result of dereddening the
photometry and a 17500 K KC model fit to the data.  The
luminosities are then calculated as a sum of the scaled KC model flux multiplied through by the distances
found previously. Taking into account the errors on the distance and temperature, the luminosities are calculated to be log(L$_*$/L$_\odot$)= 4.39 $\pm$ 0.40 for PDS 27 and log(L$_*$/L$_\odot$)= 4.75 $\pm$ 0.39 for PDS 37.

\subsubsection{Location in the HR diagram}

\begin{figure}

\includegraphics[trim=0 0 0 0 0, width=0.5\textwidth, clip=true]{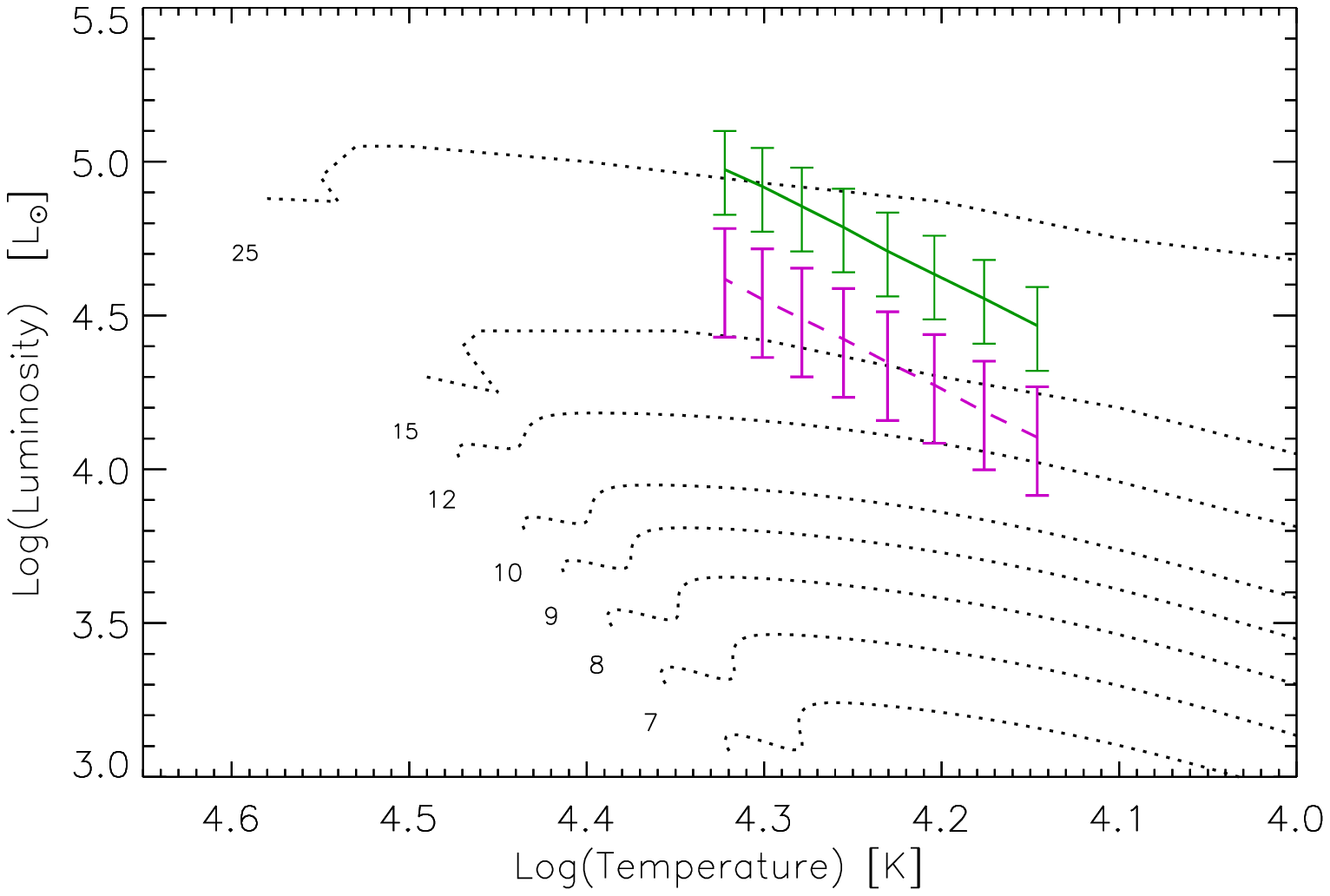}
\caption{Shows the PMS tracks of \citet{1996A&A...307..829B} and \citet{Siess2000}, along with the positions of PDS 27 in magenta dashed line and PDS 37 in green solid line. The dotted lines show where the objects lie for the determined temperature range, while the error bars show the allowed changes in luminosity given the errors in distance.}
\label{fig:pms_tracks}

\end{figure}

With the luminosity established for each star, along with a well-defined
temperature limit, the calculation of remaining stellar parameters
through the use of PMS tracks can be performed. In this work, two sets of PMS
tracks are adopted: one set by \citet{Siess2000}, as they cover up to 12~M$_\odot$, and the other set from \citet{1996A&A...307..829B}, as they cover a
larger luminosity and mass range (these are the 15 and 25~$M_\odot$ tracks in Fig. \ref{fig:pms_tracks}). Along each mass track, there is a
unique relation between the luminosity and temperature. Fig. \ref{fig:pms_tracks} shows where PDS 27 and
PDS 37 lie on the PMS tracks. Based on their position, we find a mass
of 15.3 (+5.4, -4.4)~M$_{\odot}$ for PDS 27 and a mass of 21.1 (+11.0, -5.3)~M$_{\odot}$ for PDS 37.
The radius is found straightforwardly from the {\it D/R} values, determined in the previous section, by using the distance. This provides 17.0 $\pm$ 4.0~${\rm R}_\odot$ for PDS 27 and 25.8 $\pm$ 5.0~${\rm R}_\odot$ for PDS 37. With a mass and radius determined, the log(g) is calculated to be 3.16 $\pm$ 0.27 for PDS 27 and 2.94 $\pm$ 0.35 for PDS 37. The errors in log(g) are dominated by the errors in distance. The position of the stars in the HR diagram suggests that they are very young objects, which may be evolving into
massive O-type stars.

\begin{center}
\begin{table*}

\caption{Lines tracing accretion. Errors in the EW measurements are typically 5\%. Errors in the line flux are $\sim$80\%, for both stars, due to the large temperature and distance errors assigned. EW$_{obs}$, EW$_{int}$ and EW$_{cor}$ are observed, intrinsic photospheric absorption and corrected EW, respectively.}
\label{tab:macc_lines}
{\small
\hfill{}
\begin{tabular}{ccc|ccccccc}
\hline
Line    &       Star    &       Spectra &       EW$_{obs}$      &       EW$_{int}$      &       EW$_{cor}$      &       F$_{line}$      &       log(L$_{line}$) &       log($L_{acc}$)  &       log(M$_{acc}$)  \\
        &               &               &       (${\rm \AA}$)   &       (${\rm \AA}$)   &       (${\rm \AA}$)   &       (${\rm W/m^2}$) &       (L$_\odot$)     &       (L$_\odot$)     &    (M$_\odot$/yr)     \\
\hline
H$\alpha$     &       PDS 27  &       X-shooter(a)     &       -75.20    &       4.20    &       -79.40   &       1.1$\times 10^{-13}$            &       1.54 $\pm$0.10   &       3.96$\pm$0.07   &       -3.49$\pm$0.20   \\[2pt]
H$\alpha$        &       PDS 27  &       X-shooter(b)     &       -70.40    &       4.20    &       -74.60   &       1.0$\times 10^{-13}$         &       1.52$\pm$0.10   &       3.93$\pm$0.07    &       -3.52$\pm$0.20   \\[2pt]
H$\alpha$        &       PDS 27  &       FORS2    &       -120.80   &       4.20    &       -125.24  &       1.8$\times 10^{-13}$                 &       1.74$\pm$0.10    &       4.18$\pm$0.06   &       -3.27$\pm$0.19   \\[5pt]
$[{\rm O I}]_{6300}$   &       PDS 27  &      X-shooter(a)      &       -0.28    &       0.00       &       -0.28    &       4.6$\times 10^{-16}$    &       -0.84$\pm$0.10   &       3.85$\pm$0.13    &       -3.60$\pm$0.23  \\[2pt]
$[{\rm O I}]_{6300}$       &       PDS 27  &      X-shooter(b)       &       -0.29    &       0.00       &       -0.29    &       4.8$\times 10^{-16}$    &       -0.82$\pm$0.10   &       3.87$\pm$0.14    &       -3.58$\pm$0.23   \\[2pt]
 $[{\rm O I}]_{6300}$      &       PDS 27  &       FORS2    &       -0.36    &       0.00      &       -0.36    &       5.9$\times 10^{-16}$             &       -0.73$\pm$0.10   &       3.97$\pm$0.16    &       -3.48$\pm$0.24   \\[5pt]
Br$\gamma$     &       PDS 27  &      X-shooter(a)      &       -3.80     &       4.72    &       -4.22   &       8.3$\times 10^{-16}$                &       -0.58$\pm$0.10   &       3.02$\pm$0.16    &       -4.43$\pm$0.24   \\[2pt]
Br$\gamma$       &       PDS 27  &       X-shooter(b)    &       -3.80     &       4.72    &       -4.22    &       8.3$\times 10^{-16}$               &       -0.58$\pm$0.10   &       3.02$\pm$0.16  &       -4.43$\pm$0.24   \\[5pt]

 H$\alpha$     &       PDS 37  &      X-shooter      &       -111.40   &       4.13    &       -115.53  &       2.0$\times 10^{-13}$    &       2.07$\pm$0.08    &       4.54$\pm$0.04    &       -2.87$\pm$0.22   \\[2pt]
 H$\alpha$       &       PDS 37  &       FORS2    &       -122.60   &       4.13    &       -126.73  &       2.2$\times 10^{-13}$   &       2.11 $\pm$0.08   &       4.58$\pm$0.04    &       -2.83$\pm$0.22   \\[5pt]
$[{\rm O I}]_{6300}$   &       PDS 37  &      X-shooter       &       -0.47    &       0.00       &       -0.47    &       9.6$\times 10^{-16}$    &       -0.25$\pm$0.08   &       4.51$\pm$0.36      &       -2.89$\pm$0.42   \\[2pt]
 $[{\rm O I}]_{6300}$       &       PDS 37  &       FORS2    &       -0.46    &       0.00       &       -0.46    &       9.4$\times 10^{-16}$    &       -0.26$\pm$0.08   &       4.50$\pm$0.35     &       -2.91$\pm$0.41    \\[5pt]
Br$\gamma$      &       PDS 37  &      X-shooter      &       -8.50     &       4.64    &       -9.28    &       1.2$\times 10^{-15}$    &       -0.15$\pm$0.08   &       3.41$\pm$0.49     &       -4.00$\pm$0.53

\end{tabular}}
\hfill{}
\end{table*}
\end{center}


\begin{figure*}
        \centering
        \includegraphics[width = 18cm]{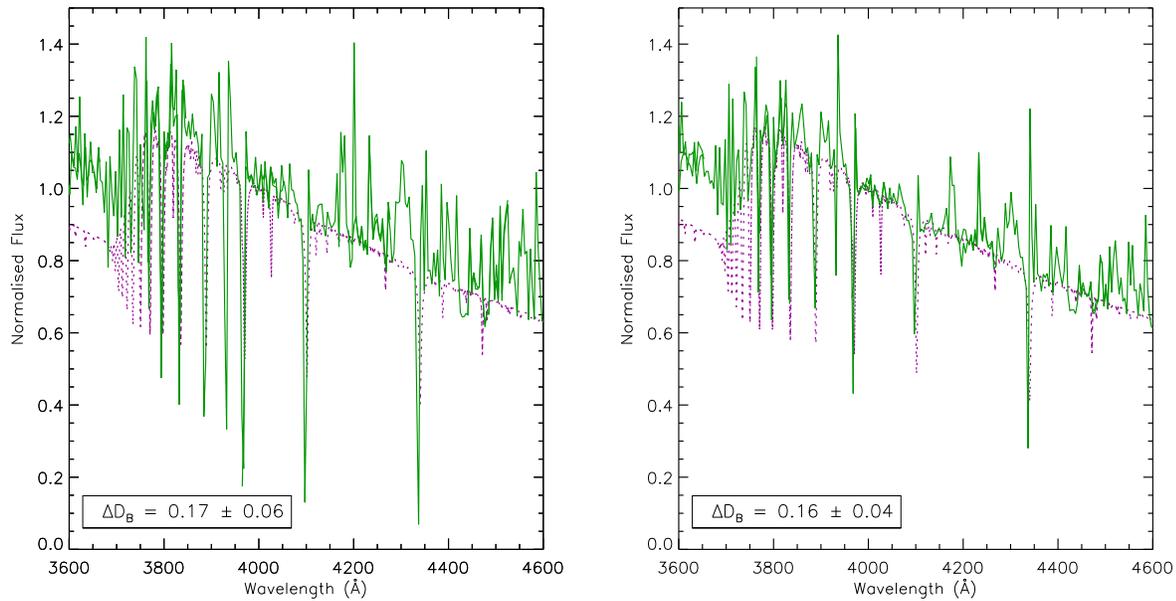}
          \caption{The Balmer jump region for PDS 27 (left) and PDS 37 (right). The targets are shown as the green solid line while the magenta dotted line is a KC model of 17500~K. The plot also highlights the extreme emission line nature of the two objects. The error in the Balmer excess displayed is solely the error due to the measurement between the two spectra.}
          \label{fig:pds37_db}
       \end{figure*}


\begin{figure}

\includegraphics[trim=0 0 0 0 0, width=0.5\textwidth, clip=true]{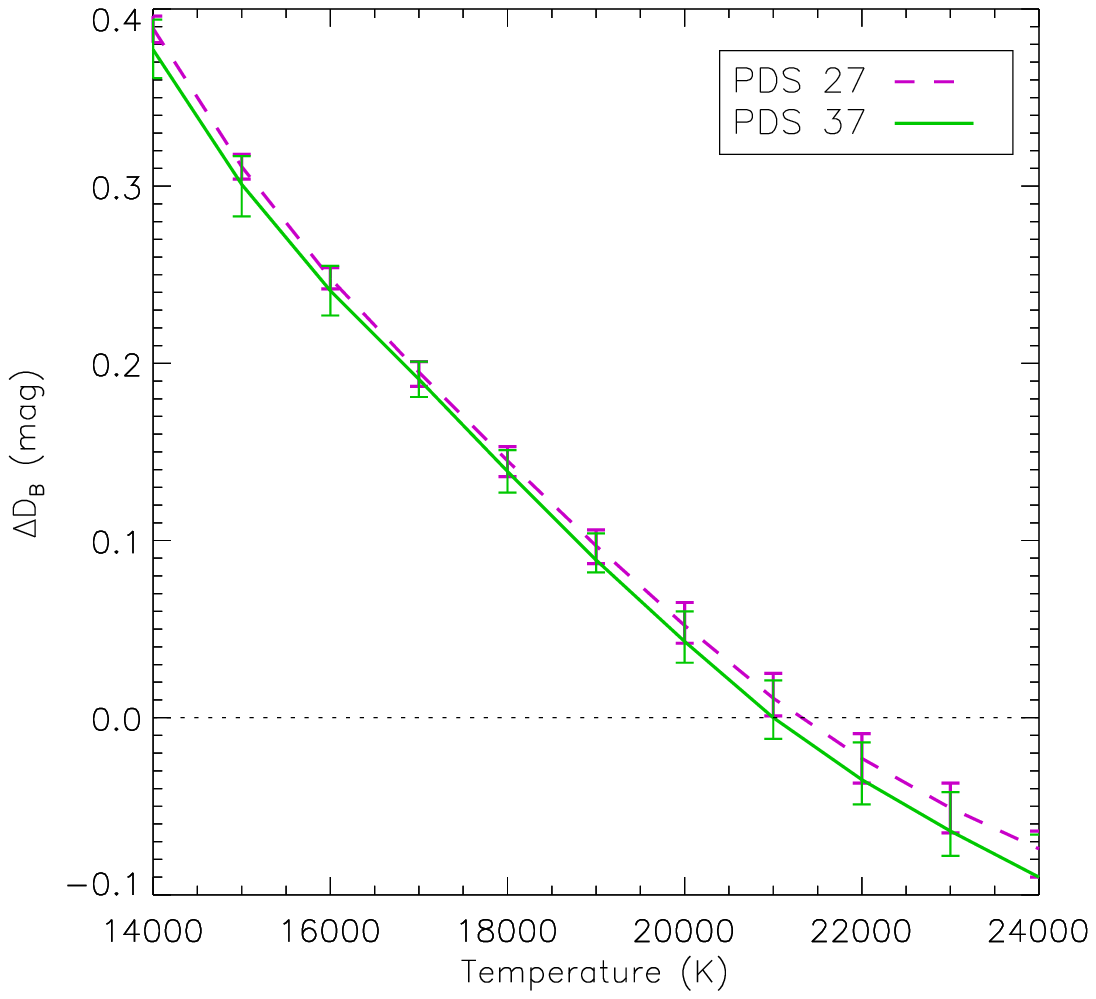}
\caption{The measured Balmer excess changes for each star
  depending upon the adopted temperature. PDS 27 is shown in magenta dashed line, and PDS 37 in green solid line. For both stars, a zero in
  Balmer excess is achieved around 21000~K.}
\label{fig:teff_db_upper}

\end{figure}

\subsection{Accretion rates}

Measurement of the accretion rates is done by two separate methods: (1) a direct relationship to the measured Balmer excess and (2) using the strength of accretion-tracing lines. 

\subsubsection{Accretion rates 1: Balmer excess}

The first method of obtaining accretion rates is to measure the
Balmer excess, $\Delta D_B$ .  This is defined as the excess of energy seen across the
Balmer jump region, and it has been observed in numerous CTTs
\citep{Gullbring00, Herczeg08, Ingleby13} and HAeBe stars
\citep{1978ApJ...224..535G, Muzerolle04, Donehew11, Mendigutia11,
  Mendigutia14}. The excess energy is often explained to be caused by magnetospheric accretion (MA), where disc material is funnelled via accretion columns on to the star where they shock the photosphere, with the final
result of this being a hotspot of increased temperature that peaks in
the UV \citep{Calvet98}. Therefore, $\Delta D_B$ serves as a direct tracer of the accretion rate.

The method of measurement here follows \citet{Donehew11}, where the
observed spectra and a model atmosphere of the star's temperature are
normalised to 4000~$\rm \AA$, and then the spectra are adjusted by a
reddening law until the slope of the SED matches that of the model
between 4000 and 4600~$\rm\AA$; this accounts for any reddening in the
spectra, and has been shown to be reddening independent
\citep{Muzerolle04,Donehew11}. Finally, the Balmer excess is measured
by the equation:

\begin{equation}
\Delta D_B= 2.5{\rm log} \left( \frac{F_*+F_{\rm acc}}{F_*} \right)
\end{equation}
where $F_*$ is the flux of the photosphere at 3600~$\rm \AA$; $F_{\rm
  acc}$ is the flux produced due to accretion, at 3600~$\rm \AA$, such
that $F_*+F_{\rm acc}$ is the observed flux. 
Fig. \ref{fig:pds37_db} shows the Balmer jump region of each object 
and the expected intrinsic photosphere, given by a KC model. The spectra is binned to 5~$\rm \AA$ resolution in order to clearly see and measure the excess, and the region measured is from 3605 to 3665~$\rm \AA$ to avoid the poor SNR region where the echelle orders crossover. $\Delta D_B$ 
is measured to be 0.17 $\pm$ 0.06 for PDS 27 and 0.16 $\pm$ 0.04 for PDS 37. However, because of the temperature range adopted, the measured $\Delta D_B$ can vary by more than this error (see Fig. \ref{fig:teff_db_upper}). Therefore, we assign larger, more generous, errors to these measurements of 0.17 (+0.13, -0.16) for PDS 27 and 0.16 (+0.22, -0.15) for PDS 37.

As mentioned in Section 4.1.3, an upper limit was placed on the temperature of the stars by considering the observed size of the Balmer jump region. For the case of stars with an excess flux in this region, due to accretion, we can instead measure the Balmer excess, which is the amount the Balmer jump is filled in by the excess flux. In Fig. \ref{fig:teff_db_upper} the observed Balmer excess for each star is shown for different temperatures (i.e. different Balmer jump sizes) to highlight that there is a limit in temperature beyond which no excess can be observed.

The relationship between $\Delta D_B$ and $\dot{M}_{\rm acc}$ depends
upon the stellar parameters of the star, and a full description is
given in \citet{Mendigutia11}. The basics are that the derivation of
$\dot{M}_{\rm acc}$ from $\Delta D_B$ is unique for each combination
of stellar parameters, i.e. in a 20000~K star the same measured
$\Delta D_B$ will yield a much higher accretion rate than a 10000~K
star because it requires a much higher accretion flux for the same
visible $\Delta D_B$. Because of this, it also means that for hotter objects smaller $\Delta D_B$ values are expected due to the sheer amount of accretion flux required to be noticeable above the photospheric level.

For the parameters determined in this work, the log($\dot{M}_{\rm
  acc}$) found are -3.96 (+0.76, -1.32) for PDS 27 and -3.56 (+0.60, -1.62) for PDS 37. Filling factors are also determined of f~=~0.40 (+0.55, -0.40) and 0.40 (0.54, -0.39), for PDS 27 and 37, respectively, when using a
typical inward flux of accretion column energy of
$10^{12}$~erg/cm$^{2}$/s \citep{Mendigutia11}. These are large but acceptable filling factors.

\subsubsection{Accretion rates 2: line luminosity }

For the second method, there are established relationships between
accretion rate and emission line strength. This is because during the
accretion process large amounts of energy are released which can
excite the circumstellar environment. This excitation and excess of
emission in CTTs can be explained by MA
\citep*{Calvet98,1998ApJ...492..743M}. Well-established line strength
relationships have been found for brown dwarfs \citep{Natta04,
  Natta06, White03}, CTTs
(\citealt*{1998AJ....116.2965M}; \citealt{Dahm08}; \citealt{Ingleby13}; \citealt{Alcala14}) and HAeBe stars
\citep{Garcia-Lopez06,Mendigutia11}. Since the two objects in this
work are early-type Herbig stars, the relationships of
\citet{Mendigutia11} are adopted. However, it should be noted that these relationships may not hold for the targets in this work which are highly luminous compared to the stars from which the relationship was derived. This is an aspect we aim to test with the direct $\Delta D_B$ derived accretion rate.

To obtain the line luminosities, a dereddened line flux is needed. This
requires a line strength to be measured, which must be corrected for intrinsic line strength. The observed, intrinsic and corrected line strengths are given in Table~\ref{tab:macc_lines}. For the Br$\gamma$ line, it should be noted that the intrinsic correction must also take into account the IR excess in the $K$ band. This is done following the steps in \citet{Garcia-Lopez06}. The flux density is then calculated from an interpolation
between the two closest photometric points; with the exception of the
Br$\gamma$ line for which no interpolation is made, instead the value of
the $K$-band magnitude is used. The photometry is dereddened by the
$A_V$ values found in Section 4.1.4 prior to this step. The line flux is obtained by multiplication of the interpolated flux density and the line
strength, and the line luminosity is then achieved  by multiplying the line flux by $4\pi D^2$.

Once the line luminosity is established, the accretion luminosity for
each line is extracted via the relationships of \citet{Mendigutia11}
which take the form of log($L_{\rm acc}$)= {\it A} + {\it B} log(L$_{\rm line}$),
where {\it A} and {\it B} are constants specific to each line. Finally, the
accretion rate is calculated via the equation: $ \dot{M}_{\rm acc} = L_{\rm acc} R_\star / GM_\star$, where {\it G} is the gravitational constant
and $\dot{M}_{\rm acc}$ is the accretion rate. Table
\ref{tab:macc_lines} gives details of the values used to derive the
accretion rates. The accretion rates derived from the H$\alpha$, [O I] and Br$\gamma$ lines are in the range of $\dot{M}_{\rm acc}\sim 10^{-3}$--$10^{-4.5}$~$M_\odot/$yr.

\section{Discussion}

We have presented a spectroscopic and spectropolarimetric study of two
early-type Herbig stars, and found that the spectra of both objects
are remarkably similar, strongly suggesting that both are in the same
evolutionary phase. The spectra are dominated by strong emission
lines, and do not reveal any photospheric absorption lines. Indeed,
the only lines that are fully in absorption are the interstellar DIB
lines.

The picture that emerges is that both objects undergo strong winds as
attested by the clear, and numerous, P Cygni profiles observed. P Cygni absorption extends in both stars to hundreds of kms$^{-1}$. The He I line at
10830~$\rm\AA$ is the only helium line seen as a P Cygni profile. The remaining
He I lines at 5875, 6678 and 7065~$\rm\AA$ are present only in absorption
(see Fig. \ref{hevari}). These lines are broader than the expected
intrinsic lines from the photosphere. If they originate from the wind,
they must probe the very hot inner wind.

\subsection{Circumstellar geometry}

The majority of the observed Fe II lines are double peaked in PDS 37,
while they are single peaked in PDS 27 (see Section 3.1). This
suggests that PDS 37 has an inclined disc that we are viewing close to
edge-on. This interpretation is supported by a high inclination angle being reported by
\citet{2013MNRAS.429.2960I}, based on first overtone CO emission. In addition, the continuum polarisation
shows a change in position angle with wavelength.  The observed
polarisation must have an intrinsic component as normally the
interstellar polarisation angle is constant with wavelength.  This
change is stronger in PDS 37 than PDS 27 as illustrated in Fig.
\ref{ser}. Further information on the geometry of the circumstellar
environment can be achieved by performing spectropolarimetry across
the lines. 

Spectropolarimetry is very effective in revealing
asymmetries in the scattering material. This technique is based on the
concept that  free electrons very close to the stars, where the
density of free electrons is high due to hydrogen ionisation, scatter
the continuum radiation from the stellar photosphere and polarise
it. The electron scattering region is of  the order of a few stellar radii \citep{1987ApJ...317..290C}. In the case of a face-on disc, or if surrounding material is
spherically symmetric, no net polarisation can be detected, while in
the case of inclined flattened structures a net polarisation can be
observed. The emission lines are less polarised than the continuum
radiation because they face fewer electrons than the continuum and the
ionised region is larger than the electron scattering region. \citet{1974MNRAS.167P..27C} and \citet{1976ApJ...206..182P} first used
spectropolarimetry in a study of Be stars, which allowed them to infer
that these stars are surrounded by discs.  

A change in polarisation across the absorptive
component of the emission line can not be explained by the above concept,
and is often suggested to be the result of the McLean
effect. In brief, the absorption takes the direct, unscattered, light
from the star away from the beam; however, the flux does not necessarily reach zero.
Some of the light emitted in other directions can be scattered into
our line of sight by free electrons, provided the electrons are in a non-spherical distribution. Therefore, the observed light in
the absorption region will be more polarised than the continuum
\citep{McLean1979}. There are many objects reported as having a
McLean effect; see for example \citet{2005MNRAS.359.1049V}

Both PDS 27 and PDS 37 show a strong McLean-type effect across the
absorptive component of H$\alpha$ (see Fig. \ref{pds}) and the
change is as broad as the absorption. This indicates the presence of
intrinsic polarisation which in turn implies that the ionised material
around the star is asymmetrically distributed. The strength of the line
effect in H$\alpha$ is of the order of 1\% in PDS 27 and 1.6\% in
PDS 37. This supports the ideas that PDS 37 has a more
inclined disc than PDS 27, as we expect more intrinsic polarisation in
a more inclined disc. The slope of the loop from the continuum to the line in the ({\it Q,U})
diagram represents the intrinsic polarisation for the continuum
polarisation (see Fig. \ref{pds}). The intrinsic continuum polarisation angle represents the electron scattering angle in the circumstellar medium. We determine an intrinsic polarisation angle of
$\sim$~$77^{\circ}$ for PDS 27 and $\sim$~$56^{\circ}$ for PDS 37. The angle has been measured from the continuum to the line in the ({\it Q,U}) diagram in Fig. \ref{pds} [see the discussion in \citet{Wheelwright11}].

We would expect a similar line effect in H$\beta$ as in H$\alpha$,
since it arises from approximately the same volume. However, we do not
see any changes across the line profile, neither in absorption nor in
the emission (see Fig. \ref{pds}). A point noteworthy is that
the absorption component in H$\beta$ is deeper than that of H$\alpha$, which is not intrinsically expected as
 the latter has a greater optical depth than the former.

In the present cases, we are able to explain this relatively
straightforwardly, by using the information that the strong emission from
H$\alpha$ is scattered into our line of sight, filling in its P Cygni
absorption component with scattered, polarised emission. In contrast,
H$\beta$ is a much weaker emission line, and therefore far fewer
photons are available to be scattered into the line of sight compared
to H$\alpha$. As a consequence, weaker, if any, changes in
polarisation across the P Cygni absorption would be predicted in this
model. This explains a non-detection of the line effect across the
absorption component in H$\beta$. This argument can also explain why
the H$\alpha$ line effect in PDS 37 is more obvious than the one in
PDS 27, as the latter line has deeper absorption components. For all of the other emission lines covered by the FORS2
spectropolarimetry of both stars, we do not detect a line effect. 

To summarise, we do not see a line effect across the H$\beta$ line as the
majority of the photons in the absorption component are unscattered
light. The observed He I emission lines are weak and they do not show any line
effect, and the line effect across the Ca II triplet in PDS 37 is not
compelling. However, from the characteristics of the observed spectral
lines, along with the continuum polarisation and the line
spectropolarimetry across the H$\alpha$ line, we can confirm that these two
objects have an asymmetric circumstellar environment. We identify this to be a circumstellar disc, whose size is at least of the order of several stellar radii.

\subsection{Stellar properties}

Based on their position in the HR diagram, we can infer that both of
the targets are young objects with an active circumstellar
environment. This is evidenced by high accretion rates of $(\dot{M}_{\rm acc})=10^{-3}$--$10^{-4.5}$~M$_\odot$yr$^{-1}$; strong winds, with P Cygni
absorption of around 90 and 80\% of continuum level for PDS 27 and
PDS 37, respectively, in H$\beta$, and also maximum respective
velocities of $\sim$-700 and $\sim$-400~kms$^{-1}$ measured in the H$\alpha$ line; and variability in line strengths and velocities seen in both the hydrogen and the He I lines. In
particular, PDS 27 has shown unusual variability in the He I lines
where the velocity of the line centre and the line strength change dramatically between epochs. This type of variability in line strength has
been observed previously in other objects by
\citet{2013A&A...554A.143K}, but an origin of the varying line strength cannot be reached with only three epochs of data. PDS 27 also shows variability in the observed {\it V$_{\rm LSR}$} between the X-shooter and FORS2 spectra. This could be due to the presence of a secondary companion. However, with only three epochs of data, we cannot establish conclusively the binary nature of PDS 27; additional data are required to determine the frequency and magnitude of the velocity variations. Unlike PDS 27, PDS 37 did not show any variability in the observed {\it V$_{\rm LSR}$} between X-shooter and FORS2 spectra.

Their young age is further supported by their large radii and low surface
gravities. These properties are typical of very young objects where
their puffed-up envelope is still undergoing contraction and will
shrink as it progresses towards the main-sequence
\citep{Davies2010}. The accretion rates obtained from the H$\alpha$ and [O I] 6300~$\rm\AA$ are found to be the same across all epochs, within the errors, for both PDS 27 and PDS 37. These two lines also show that PDS 37 has an accretion rate $\sim$0.6 dex greater than PDS 27, which is comparable to the difference seen between the accretion rates derived from the Balmer excess of $\sim$0.4; both are within the errors. The Br$\gamma$ line does not follow the same trends as it shows a decrease in accretion rate, in both objects, over both the other two lines and the Balmer excess measurement (see Table \ref{tab:macc_lines} for all of the emission line derived accretion rates).

The accretion rates derived using the Br$\gamma$ line are higher than
any other reported accretion rates in HBe stars, which also use the Br$\gamma$
line, by $\sim$1-3 orders of magnitude
\citep{Garcia-Lopez06, Donehew11}. The Balmer excess method results in 
${\rm log}(\dot{M}_{\rm acc})=$-3.96 (+0.76, -1.32) for PDS 27 and -3.56 (+0.60, -1.62) for PDS
37. These values agree with the respective accretion rates derived
from all of the lines analysed. The Br$\gamma$ line in both objects shows a difference of an order of magnitude less in accretion rate between the other lines.
A possible breakdown in the relationship between line luminosity and accretion luminosity has been suggested by \citet{Donehew11}, where the luminosity increases towards the early-type
HAeBe star regime. However, the two stars presented in this work still follow this relationship. We cannot conclude with just these two objects whether the relationship holds or not, and such is
beyond the scope of this paper. Instead, PDS 27 and PDS 37 will form
part of an upcoming paper exploring these relationships in a larger
sample (Fairlamb et al., in preparation).

\section{Conclusions}

This work presents new data on two previously poorly studied HAeBe
candidate stars. By combining X-shooter spectroscopy and FORS2
spectropolarimetry, our understanding of properties in the
circumstellar environments and the stellar parameters of the stars themselves has
improved. The main findings in this work are as follows:

\begin{itemize}

\item The targets are both very hot 17500 $\pm$ 3500~K with a spectral type
  of around B2. When combined with their high luminosities, of log(L$_*$/L$_\odot) \sim 4.5$, they are placed in a secluded area on the
  HR diagram where very young objects lie. They are swollen objects with large
  radii, 17.0 $\pm$ 4.0 and 25.8 $\pm$ 5.0~${\rm R}_\odot$, and large masses, 15.3 (+5.4, -4.4) and 21.1 (+11.0, -5.3)~M$_{\odot}$, for PDS
  27 and 37 respectively. Their young nature is supported by the
  high accretion rates calculated here, which are among the highest measured in
  HAeBe stars to date, $\dot{M}_{\rm acc}\sim 10^{-3}$--$10^{-4.5}$~M$_\odot$yr$^{-1}$.

 \item Spectropolarimetric line effects are detected in the H$\alpha$ line for
   both objects. This polarisation property is explained by the
   McLean effect; line photons are scattered into the line of sight,
   increasing the polarisation across the P Cygni absorption. The
   spectropolarimetry indicates that the circumstellar environment is
   of a flattened structure, likely a disc.

\item No line effects are seen across lines other than H$\alpha$. This
  is consistent with the notion that the spectropolarimetric behaviour
  in both cases is due to the McLean effect. The effect requires a strong
  emission line filling in absorption and a line with great optical depth so that the
  scattered photons will dominate the polarisation. It appears that in
  the cases studied here, only H$\alpha$ fulfils both requirements.

\item Line variability is seen in both objects, which we ascribe to an
  active circumstellar environment. Extreme line variability of the He
  I species is seen in PDS 27. In addition, PDS 27 also shows a variability in the observed {\it V$_{\rm LSR}$} between X-shooter and FORS2; this could be due to the fact that the system is binary.

\end{itemize}

Finally, we have presented two young PMS stars, whose
spectral classification as early B-type is consistent with them belonging
to the group of HAeBe stars. Their masses and locations in the
HR diagram indicate that they will continue to evolve to
become O-type stars. As they are optically bright, these stars provide
important test beds to study the formation of the most massive stars.

\section*{Acknowledgements}

The authors would like to thank the ESO staff for carrying out the
observations in service mode. Jorick Vink is thanked for helpful discussions. KMA is supported by the Human Capacity
Development Program (HCDP) in
Kurdistan-Iraq. JRF is supported by the UK Science and Technology
Facilities Council (STFC).

\bibliographystyle{mn2e}

\label{lastpage}
\end{document}